\documentclass[5p,times]{elsarticle}

\usepackage{lineno,hyperref}
\modulolinenumbers[5]

\journal{Journal of \LaTeX\ Templates}

\usepackage{amsmath,amssymb,setspace,graphicx,subfigure,epsfig}
\usepackage{amsfonts}
\usepackage{latexsym}
\usepackage{epsfig}
\usepackage{color}
\usepackage{enumerate}% http://ctan.org/pkg/enumerate
\usepackage{caption}

\newcommand{\dis}[1]{\begin{equation}\begin{split}#1\end{split}\end{equation}}
\newcommand{\be}{\begin{equation}}
\newcommand{\ee}{\end{equation}}
\def\bea{\begin{eqnarray}}
\def\eea{\end{eqnarray}}
\newcommand\ba{\begin{eqnarray}}
\newcommand\ea{\end{eqnarray}}

\newcommand{\eq}[1]{Eq.~(\ref{#1})}

\newcommand\gev{\,{\rm GeV}}
\newcommand\mev{\,{\rm MeV}}
\newcommand\kev{\,{\rm keV}}
\newcommand\ev{\,{\rm eV}}

\newcommand\cm{\,{\rm cm}}

\begin{document}

\begin{frontmatter}

\title{Constraining dark photon model with dark matter from CMB spectral distortions}

\author[CNU]{Ki-Young Choi}
\ead{kiyoungchoi@jnu.ac.kr}

\author[CTPU]{Kenji Kadota}
\ead{kadota@ibs.re.kr}

\author[KAIST,CTPU]{Inwoo Park\corref{mycorrespondingauthor}}
\cortext[mycorrespondingauthor]{Corresponding author}
\ead{inwpark@kaist.ac.kr}

\address[CNU]{Institute for Universe and Elementary Particles and Department of Physics, Chonnam National University, 77 Yongbong-ro, Buk-gu, Gwangju, 61186, Republic of Korea}
\address[CTPU]{Center for Theoretical Physics of the Universe,
Institute for Basic Science (IBS), Daejeon, 34051, Korea}
\address[KAIST]{Department of Physics, Korea Advanced Institute of Science and Technology, Daejeon, 34141,Republic of Korea}

\begin{abstract}
  Many extensions of Standard Model (SM) include a dark sector which can interact with the SM sector via a light mediator. We explore the possibilities to probe such a dark sector by studying the distortion of the CMB spectrum from the blackbody shape due to the elastic scatterings between the dark matter and baryons through a hidden light mediator. We in particular focus on the model where the dark sector gauge boson kinetically mixes with the SM and present the future experimental prospect for a PIXIE-like experiment along with its comparison to the existing bounds from complementary terrestrial experiments.
  
\end{abstract}


\end{frontmatter}

%\linenumbers

\section{Introduction}
\label{intro}
The energy spectrum of the cosmic microwave background (CMB) follows the most perfect blackbody spectrum ever observed. There yet can exist a minuscule deviation from the blackbody when the CMB photons are not in a perfect equilibrium.
The number-changing interactions such as Bremsstrahlung and double Compton scatterings are not efficient enough for the redshift $z\lesssim 2\times 10^6$ and the energy injection/extraction can result in the Bose-Einstein distribution with a non-vanishing $\mu$ parameter (rather than the blackbody distribution with $\mu=0$)~\cite{Sunyaev:1970er}. For $z\lesssim 5\times 10^4$, even the kinetic equilibrium cannot be maintained due to the inefficient Compton scatterings and the spectrum distortion can be characterized by the Compton $y$-parameter which is given by the line of sight integral of electron pressure~\cite{Zeldovich:1969ff}.

The attempt to measure potential CMB spectral distortion has been made by the Far Infrared Absolute Spectrophotometer (FIRAS) instrument aboard the COBE satellite~\cite{Fixsen:1996nj} two decades ago, leading to the upper bounds $|\mu| \lesssim 10^{-4}$ and $|y|\lesssim 10^{-5}$. The next generation space-telescope PIXIE~\cite{Kogut:2011xw} is expected to improve the sensitivity to $|\mu|\sim 5\times 10^{-8}$ and $|y|\sim 10^{-8}$. 

The CMB spectral distortion can, for instance, be induced by the energy injection into the background plasma in many non-standard cosmological scenarios~\cite{Chluba:2011hw}. The examples include the energy release from decaying heavy relics~\cite{Hu:1993gc,Sarkar:1984tt}, evaporating primordial black holes~\cite{Carr:2009jm}, the annihilating dark matter (DM)~\cite{McDonald:2000bk,Padmanabhan:2005es} and the dissipation of acoustic waves~\cite{Sunyaev:1970,Chluba:2012gq,Silk:1967kq}.

Even in the standard cosmology, however, the CMB distortion can occur due to the energy transfer between the photons and the ``baryons'' (protons and electrons) \cite{Chluba:2011hw,Khatri:2011aj,Pajer:2013oca}. The Coulomb interactions of non-relativistic plasma consisting of baryons with photons can extract energy from the CMB and maintain the kinetic equilibrium. The temperature of baryons follows that of photons and decreases inversely proportional to the scale factor of the Universe, $T_b \simeq T_\gamma \sim 1/a$, instead of $1/a^2$ for the decoupled non-relativistic matter. This extraction of energy from the CMB results in the $\mu$-distortion of the order of $\mu\simeq -3\times 10^{-9}$.

The analogous effects can be induced when the DM is thermally coupled to the photon-baryon plasma by the elastic scatterings, and such effects on the CMB spectral distortions were first discussed in~\cite{tashiro2014}  and elaborated on in~\cite{Ali-Haimoud:2015pwa}. 
The additional energy extraction from CMB into DM  enhances the spectral distortion of CMB with a negative $\mu$. Since the DM number density is inversely proportional to its mass, for a given DM mass density, the FIRAS can constrain the DM mass up to $m_{\chi}\sim 0.1$ GeV and a future experiment such as PIXIE can further extend its sensitivity to $m_{\chi} \sim 1\gev$. The CMB distortion measurements would complement the other heavy DM searches such as the direct detection experiments which rapidly lose the sensitivity to sub-GeV DM due to the small recoil energy of the nuclear target.

One of the intriguing models which can realize the coupling of the DM to the SM particles  is a "dark photon" scenario where there exists a dark sector with a broken U(1) gauge symmetry~\cite{Okun:1982xi,Holdom:1985ag}. The phenomenology associated with such a novel dark sector has received considerable attention in recent years and a wide range of experimental searches have been performed in the collider and beam dump experiments such as BarBar, PHENIX, E137 and Charm~\cite{Davoudiasl:2012ag,Essig:2013lka,Adare:2014mgk,Goudzovski:2014rwa,Lees:2015rxq,Alekhin:2015byh}. 
  The constraints on the dark photon model from the cosmological and astrophysical observations have also been discussed recently~\cite{Dvorkin:2013cea,Berger:2016vxi}.

In this paper, we study the spectral distortion of CMB in the dark photon model, where the DM and baryons can interact via a dark photon, caused by the momentum transfer between CMB and DM via the elastic scatterings. We also illustrate the comparison with the existing constraints on the dark photon model in the laboratory and astrophysical observations. We first review the model in \S \ref{model} followed by the estimation of CMB distortions in \S \ref{cmbdmb}.
\S \ref{figsec} gives our results, followed by the conclusion in \S \ref{Conclusion}.

\section{Dark photon and DM}
\label{model}
We consider the dark sector consisting of the dark photon and DM.
We assume that $U(1)_d$ gauge symmetry in the dark sector has a kinetic mixing with $U(1)_Y$ in the SM of  $SU(3)_C\times SU(2)_L\times U(1)_Y$~\cite{Okun:1982xi, Holdom:1985ag}. The mixing is parametrised by a small parameter $\varepsilon$ as
\dis{
   \mathcal{L}_{mixing} =  \frac{\varepsilon}{2} \hat{B}_{\mu \nu}\hat{Z}_d^{\mu \nu}
   \label{mixing}
}
 where $\hat{B}_{\mu \nu}$ and $\hat{Z}_{d\mu \nu}$ are the field strengths of $U(1)_Y$ and $U(1)_d$ respectively.
We also assume that the fermion DM $\chi$ has the  $U(1)_d$ gauge interaction with the gauge coupling $g_d$ as
\dis{
  \mathcal{L}_{int} = -g_d \hat{Z}_{d\mu}\overline{\chi}\gamma^\mu\chi.
} 

 After the electroweak symmetry breaking, we replace $\hat{B}_{\mu\nu}= -s_W \hat{Z}_{\mu \nu} + c_W \hat{A}_{\mu \nu}$ with $s_W=\sin\theta_W$ and $c_W=\cos\theta_W$ and the mass of $\hat{Z}_{\mu \nu}$, $m_Z^0$, is generated from the Higgs mechanism. Similarly we assume that the hidden gauge boson has a mass $m_{Z_d}^0$ by $U(1)_d$ symmetry breaking through the hidden sector Higgs mechanism.

The kinetic mixings between the gauge fields can be removed and the kinetic terms can be canonically normalized by the following field re-definition
   \ba
\left( \begin{array}{c} A_{SM\mu} \\ \phantom{1} \\ Z_\mu^0 \\ \phantom{1} \\ Z_{d\mu}^0 \end{array} \right) = \left( \begin{array}{ccc} 1&0 & -\varepsilon c_W \\ \phantom{1}&\phantom{1} & \phantom{1} \\0&1&\varepsilon s_W \\ \phantom{1}&\phantom{1} & \phantom{1}\\ 0& 0& \sqrt{1-\varepsilon^2} \end{array} \right) \left( \begin{array}{c} \hat{A}_\mu\\ \phantom{1} \\ \hat{Z}_\mu \\ \phantom{1} \\ \hat{Z}_{d\mu} \end{array} \right)
\ea
  leading to
   \ba
\mathcal{L}&=&-\frac{1}{4}A_{SM\mu\nu}A_{SM}^{\mu\nu}-\frac{1}{4}Z^0_{\mu\nu}Z^{0\mu\nu}-\frac{1}{4}Z^0_{d\mu\nu}Z_d^{0\mu\nu}\nonumber\\
&&+\frac{1}{2}{m_Z^0}^2Z^0_\mu Z^{0\mu}-{m_Z^0}^2\frac{\varepsilon s_W}{\sqrt{1-\varepsilon^2}}Z^0_\mu {Z^0_d}^\mu\nonumber\\
&&+\frac{1}{2}\left({m_Z^0}^2\frac{\varepsilon^2s^2_W}{1-\varepsilon^2} +{m_{Z_d}^0}^2\frac{1}{1-\varepsilon^2} \right)Z^0_{d\mu}Z_d^{0\mu}
\ea

The mass matrix of $Z^0_\mu$ and $Z^0_{d\mu}$ can be diagonalised by a mixing parameter $\theta_X$,
\ba
\left( \begin{array}{c} Z_{SM\mu} \\ \phantom{1} \\ Z_{d\mu} \end{array} \right) = \left( \begin{array}{cc} \cos\theta_X & -\sin\theta_X \\ \phantom{1} & \phantom{1} \\ \sin\theta_X & \cos\theta_X \end{array} \right) \left( \begin{array}{c} Z^0_\mu \\ \phantom{1} \\ Z^0_{d\mu} \end{array} \right)
\ea
where \ba\tan2\theta_X=\frac{2{m_Z^0}^2\varepsilon s_W/\sqrt{1-\varepsilon^2}}{{m_Z^0}^2-{m_Z^0}^2\{\varepsilon^2s^2_W/(1-\varepsilon^2)\} -{m_{Z_d}^0}^2\frac{1}{1-\varepsilon^2}}.
\ea
The bare gauge fields are consequently related to the mass eigenstates as
\dis{
\hat{A}_\mu &= A_{SM\mu}-\frac{\varepsilon c_Ws_X}{\sqrt{1-\varepsilon^2}}Z_{SM\mu}+\frac{\varepsilon c_Wc_X}{\sqrt{1-\varepsilon^2}}Z_{d\mu},\\
\hat{Z}_{d\mu} &=-\frac{s_X}{\sqrt{1-\varepsilon^2}}Z_{SM\mu}+\frac{c_X}{\sqrt{1-\varepsilon^2}}Z_{d\mu},\\
\hat{Z}_{\mu} &= \left(c_X+\frac{\varepsilon s_Ws_X}{\sqrt{1-\varepsilon^2}}\right)Z_{SM\mu}+\left(s_X-\frac{\varepsilon s_Wc_X}{\sqrt{1-\varepsilon^2}}\right)Z_{d\mu},
}
where $s_X=\sin \theta_X$ and $c_X=\cos \theta_X$.

The electromagnetic current hence has the interaction
\dis{
 \mathcal{L}_{int}=-e J^{\mu}_{em}\left( A_{SM\mu} - \frac{\varepsilon c_Ws_X}{\sqrt{1-\varepsilon^2}} Z_{SM \mu}+\frac{\varepsilon c_Wc_X}{\sqrt{1-\varepsilon^2}}Z_{d\mu} \right),
} 
and the DM interacts with $Z_{d\mu}$ and $Z_{\mu}$ as
\dis{
 \mathcal{L}_{int}=-g_d\overline{\chi}\gamma^\mu\chi \left(  \frac{c_X}{\sqrt{1-\varepsilon^2}}Z_{d\mu}  -  \frac{s_X}{\sqrt{1-\varepsilon^2}}Z_{SM\mu}\right).
}

\begin{figure} [h]
\begin{center}
\includegraphics[width=7cm]{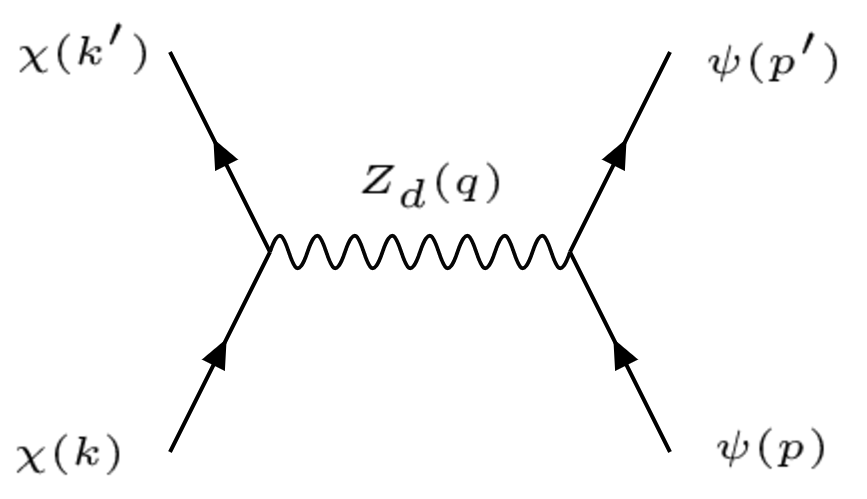}
\end{center}
\caption{Elastic scattering between baryon ($\psi$) and DM ($\chi$) through a dark photon ($Z_d$) exchange.  }
\label{diagram}
\end{figure}

We can therefore see that the electromagnetic current in the SM which couples to $\hat{A}_{\mu}$ can interact with the dark photon $Z_d$ suppressed by $\varepsilon$.
Since we are interested in the parameter range $m_{Z_d}\sim \gev  \ll m_Z$,  
we can represent our dark sector model with two free parameters $\varepsilon$ and $m_{Z_d}$ in the following sections. We hence discuss the CMB spectral distortions when the DM interactions with the SM fields $\psi_{SM}$ are mediated by the dark photon, represented by the Lagrangian
\dis{
 \mathcal{L}_{int}= -e\varepsilon c_W \overline{\psi}_{SM}\gamma^\mu\psi_{SM}  Z_{d\mu}-g_d\overline{\chi}\gamma^\mu\chi  Z_{d\mu} .
}
The corresponding Feynman diagram is shown in Fig \ref{diagram}. 
We note here that the DM does not interact with the SM photon and only couples to the SM particles by mediating $Z_d$ gauge boson \footnote{The DM coupling to the SM $Z$ is suppressed by $\tan\theta_X$ compared with that to dark photon and hence negligible in the limit of $m_{Z_d}\ll m_Z$ and a small $\varepsilon$.}.  

   %%%%%%%%%%%%%%%%%%%   
%%%%%%%%%%%%%%%%%%%%%
  
  \section{CMB spectrum distortion from DM-baryon scattering}
  \label{cmbdmb}
For the decoupled non-relativistic DM, the temperature decreases as $T_\chi \sim a^{-2}$ ($a$ is the scale factor). When DM is kinetically coupled to the background baryons ($z\gtrsim 10^4$), however, $T_\chi$ evolves along with baryon temperature $T_b$ obeying the Boltzmann equation~\cite{Dvorkin:2013cea,Ali-Haimoud:2015pwa} 
\dis{
\label{tempeq} 
\dot{T}_\chi = -2H T_\chi +\Gamma_{\chi b} (T_b-T_\chi),
}
with 
\dis{
\Gamma_{\chi b} = \frac{2c_n N_b \sigma_n m_b m_\chi}{(m_b+m_\chi)^2} \left( \frac{T_b}{m_b} +\frac{T_\chi}{m_\chi} \right)^{(n+1)/2},
\label{Gamchib}
}
where $m_b$, $N_b=N_b^0 a^{-3}$ are the baryon mass and number density. $c_n$ is a constant of the order of unity depending on the power $n$ of the DM-baryon elastic scattering cross section $\sigma_{tr} (v) =  \sigma_n v^n$ with $v$ being the DM-baryon relative velocity.  We use the conventional cross section for the momentum-transfer
   \begin{equation}
   \label{eq:sig}
\sigma_{tr}\equiv\int d\Omega(1-\cos\theta)\frac{d\sigma}{d\Omega}.
\end{equation}
   where the weight factor $(1-\cos \theta)$ represents the longitudinal momentum transfer and regulates the spurious infrared divergence for the forward scattering (corresponding to no momentum transfer with $\cos \theta \rightarrow 1)$~\cite{raby1987}.

   The DM-baryon scatterings can cause the distortion of the photon spectra and the rate of the photon energy extraction from these elastic scatterings becomes \cite{Chluba:2011hw,Ali-Haimoud:2015pwa}
   \dis{
   \label{derivrho}
   \rho_{\gamma} \frac{d}{dt}
\left( \frac{\Delta \rho_{\gamma}}{\rho_{\gamma}} \right)
= - \frac{3}{2} \left( N_b^{tot} + r_{\chi b} N_{\chi} \right) H T_{\gamma},
}
where $r_{\chi b}\equiv \Gamma_{\chi b}(T_b-T_{\chi})/(H T_b)$ parametrises  the efficiency of the momentum transfer from photons to DM, while the first term on RHS represents the energy transfer from the photons to baryons due to Compton scattering. The baryon number density $N_b^{tot}=\rho_b/m_H(2-\frac{5}{4}Y_{He})$, with $m_H$ the mass of the hydrogen, and $Y_{He}$ helium fraction by mass.
Its integration can give the estimation for the amplitude of the spectral distortion $\Delta\equiv\Delta\rho_{\gamma}/\rho_{\gamma}$. The observational bound from the FIRAS is $|\Delta| \lesssim 6 \times 10^{-5}$, and this bound is expected to be improved for the PIXIE to the level of $\Delta\approx 10^{-8}$.

   For a simple power law form of the DM-baryon elastic scattering cross section $\sigma_{tr} (v) =  \sigma_n v^n$, the FIRAS gives the upper bound on the cross section as~\cite{Ali-Haimoud:2015pwa}
\dis{
\sigma_n \leq \sigma_n^{max} \equiv C_n \frac{m_\chi}{m_b}\left( 1+\frac{m_b}{m_\chi}\right)^{\frac{3-n}{2}} \left(\frac{a_{\rm max}}{a_\mu}  \right)^{\frac{n+3}{2}\frac{m_\chi}{m_\chi^{\rm max}}},
\label{sigmax}
}
$a_{\rm max}=10^{-4}$, $a_\mu = 0.5\times 10^{-7}$ with $m_\chi^{\rm max}=0.18\mev$ (the same formulae are applicable for the future sensitivity of PIXIE with the replacement $m_\chi^{\rm max}=1.3\gev$).
For the DM-proton scattering, $C_n = (1.4\times 10^{-30}, 1.1\times 10^{-27}, 8.2\times 10^{-25}, 5.5\times 10^{-22})$ $\cm^2$ for $n=(-1,0,1,2)$ respectively  and $m_b$ with the proton mass $m_p$~\cite{Dvorkin:2013cea}.

The analogous bounds can be obtained for the scatterings between DM and electrons by replacing the coefficients $C_n$ in \eq{sigmax} with $C_n = (1.4\times 10^{-30}, 2.6\times 10^{-29}, 4.5\times 10^{-28}, 7.0\times 10^{-27})$ $\cm^2$  for $n=(-1,0,1,2)$ respectively and $m_b$ with the electron mass $m_e$.

\begin{figure} [b!]
\begin{center}
\includegraphics[width=9cm]{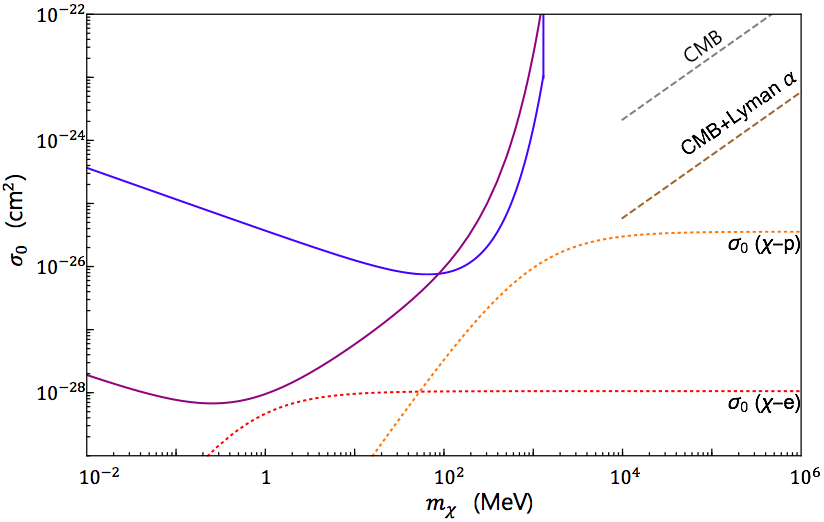}
\end{center}
\caption{The expected upper bound on the cross section from the PIXIE-like CMB spectral distortion experiment is shown with the solid lines:  $\sigma_0^{max} (\chi-p)$ for DM-proton scattering (blue) and $\sigma_0^{max} (\chi-e)$ for DM-electron scattering (purple) respectively~\cite{Ali-Haimoud:2015pwa}. We also show the constraints from Planck CMB data, and CMB+ SDSS Lyman $\alpha$ data~\cite{Dvorkin:2013cea} with dashed lines for comparison. The cross sections in the dark photon model are shown with dotted lines:  $\sigma_0 (\chi-p)$ for the interaction of DM with protons while  $\sigma_0 (\chi-e)$ with electrons. Here we used $\alpha_D=0.1$,  $m_{Z_d}=1$ MeV and $\varepsilon=10^{-5}$ (for DM-proton) and $10^{-3}$ (for DM-electron). }
\label{Bndn0}
\end{figure}

\section{CMB spectral distortion in dark photon model}
\label{figsec}
We now consider new constraints on the dark photon model from the CMB spectral distortions due to the elastic scatterings between DM and baryons. CMB distortions can probe the DM mass smaller than $\gev$ and complement the existing bounds from other experiments as we shall discuss in the following.

In the dark photon model with a kinetic mixing outlined in \S \ref{model}, the momentum transfer between DM and the baryon is mediated by the dark photon as in Fig.~\ref{diagram}.  The corresponding matrix element is
\begin{equation} \label{eq:msq}
|\mathcal{M}|^2=\frac{64\pi^2c_W^2\varepsilon^2\alpha\alpha_D}{(q^2-m_{Z_d}^2)^2}\Big[4(k\cdot p)(k'\cdot p)+m_b^2q^2+k\cdot k'q^2+q^4\Big], \end{equation}
where $\alpha\equiv e^2/4\pi \simeq 1/137$ and $\alpha_D \equiv g_d^2/4\pi$.
Here DM momentum and the relative velocity of baryon-DM in the CM frame are related as  $|\vec{k}|=v m_\chi m_b/(m_\chi+m_b)$
assuming both the baryon and DM are non-relativistic.
The corresponding momentum transfer cross section  for  $m_{Z_d}\gg|\vec{k}|$ is given by
\begin{equation}
\sigma_{tr}=\frac{16\pi c_W^2\varepsilon^2\alpha\alpha_D}{(m_\chi+m_b)^2m_{Z_d}^4}m_\chi^2 m_b^2+ O(v^2).
\end{equation}
Note that the leading term is independent of the velocity for the non-relativistic hidden gauge boson.

Fig.~\ref{Bndn0} shows how the momentum-transfer cross section varies in terms of $m_\chi$  (dotted lines) along with the expected upper bounds from the CMB distortion with the PIXIE-like sensitivity $\Delta \simeq 10^{-8}$, (solid lines). The region above $\sigma_0^{\rm max}$ is disfavored due to the large spectral distortion. For the PIXIE experiment, the constraint can be applied for the DM mass $m_\chi \le 1.3$ GeV, since, for a larger DM mass, the distortion is too small due to the smaller DM abundance as $N_\chi/N_b^{tot} \sim 3 (\gev/m_\chi)$ ~\cite{Ali-Haimoud:2015pwa}.
The dotted lines represent the constraints from the Planck CMB and SDSS Ly$\alpha$ forest data obtained in Ref.~\cite{Dvorkin:2013cea} whose analysis are applicable only to heavier DM $m_\chi \ge 10$ GeV for comparison.

%%%%%%%%%%%
\begin{figure} [t!]
\centering
\includegraphics[width=9cm]{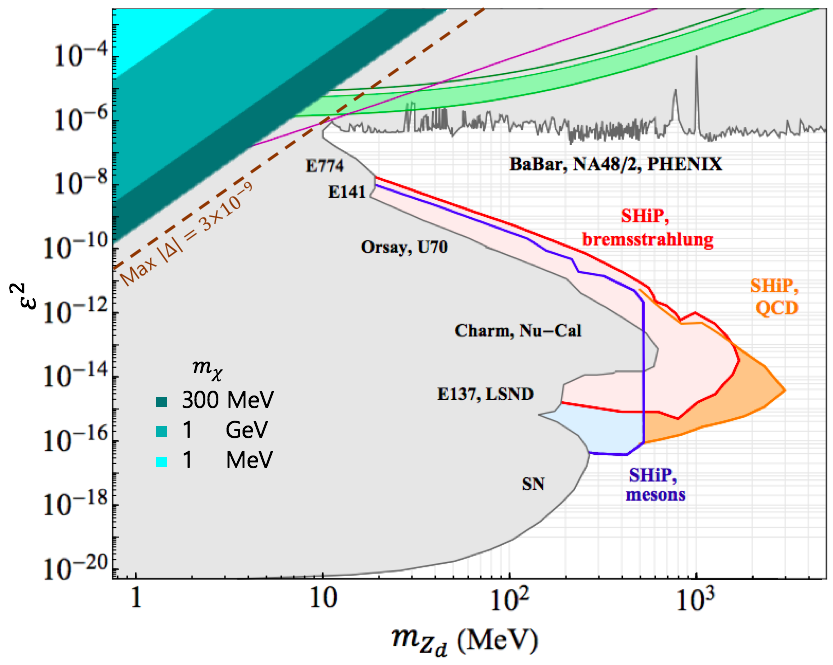}
\caption{The expected  bounds from the CMB spectral distortion by PIXIE (colored regions are excluded) when $m_{Z_d}\gg \kev$ for a few representative DM masses ($m_\chi=1 \mev, 300\mev, 1 \gev$), due to the elastic scattering between DM and protons. $\alpha_d$ = 0.1 is used for concreteness and the parameter sets producing the CMB distortion of the order $|\Delta|\approx 3 \times 10^{-9}$ expected in the conventional standard cosmology are indicated in a dashed line (brown). The other experimental constraints are adopted from~\cite{Alekhin:2015byh}.}
\label{protonDM}
\end{figure}
%%%%%%%%%%%

%%%%%%%%%%%
\begin{figure} [t!]
\centering
\includegraphics[width=9cm]{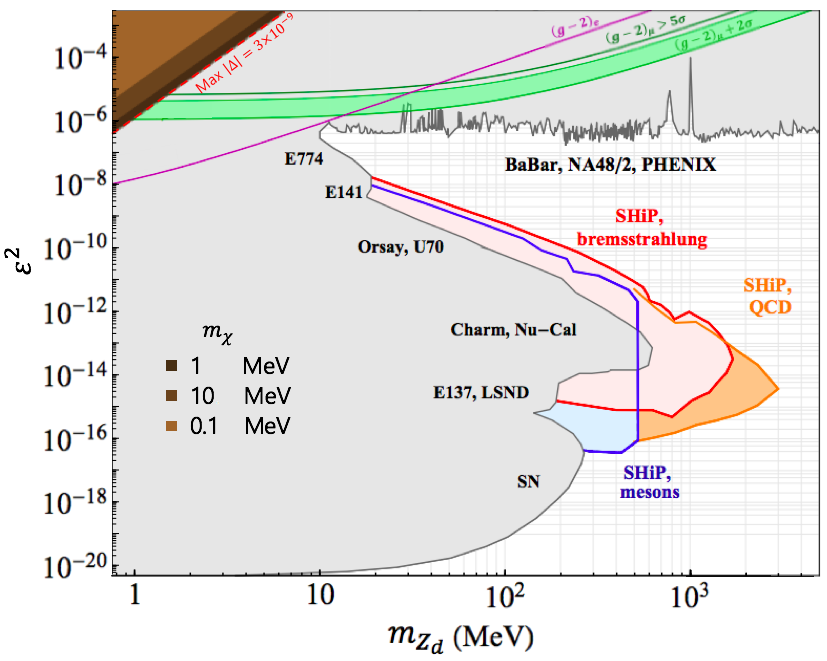}
\caption{The bounds due to the elastic scattering between DM and electrons, to be compared with the bounds from the DM-proton scattering in Fig.~\ref{protonDM}. 
%\cred{Constraints from the spectral distortion due to electron-DM scattering is weaker than those from the proton-DM scattering. }
}
 \label{electronDM}
\end{figure}
%%%%%%%%%%%

%%%%%%%%%%%
\begin{figure}[!]
\centering
\includegraphics[width=9cm]{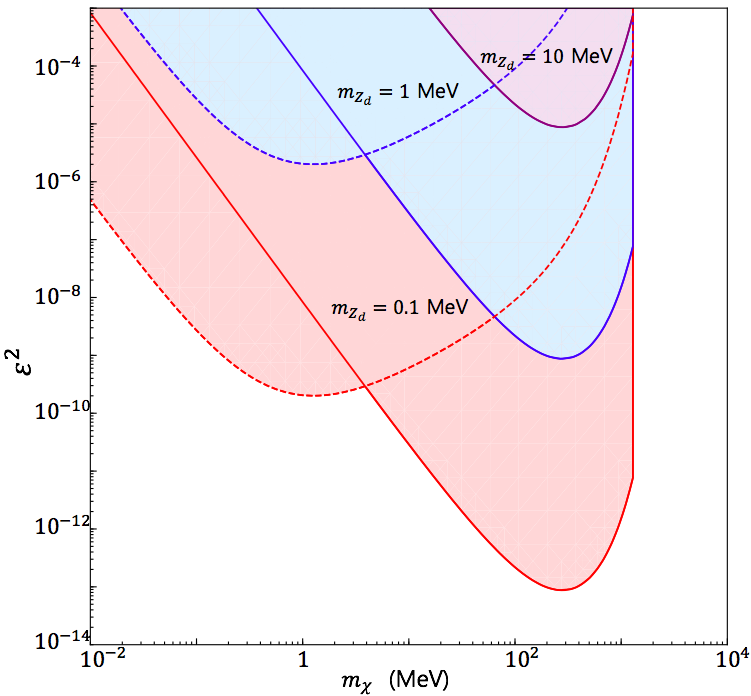}
\caption{The expected upper bounds from PIXIE (colored regions are excluded) in terms of the DM mass ($m_\chi$) and the kinetic mixing ($\varepsilon^2$) for $m_{Z_d}\gg \kev$. The bounds from the DM-proton (DM-electron) scattering are shown with solid (dashed) lines. Different colors are for a few representative dark photon masses ($m_{Z_d}=0.1 \mev,1 \mev, 10 \mev$) and $\alpha_d$ = 0.1 is used for concreteness.}
\label{fig:EpsMchiPro}
\end{figure}
%%%%%%%%%%%

Figs.~\ref{protonDM} and~\ref{electronDM} show the bounds from the CMB distortion on the dark photon mass ($m_{Z_d}$) and the kinetic mixing ($\varepsilon^2$) for different DM masses.
We show the constraints from the DM-proton interaction with $m_\chi=1\mev,300\mev,1\gev$ in Fig.~\ref{protonDM}, and those from the DM-electron interaction with $m_\chi=0.1\mev,$ $1\mev$,   $100\mev$ in Fig.~\ref{electronDM}.
   We here used $\alpha_D=0.1$ and $m_\chi^{\rm max}=1.3 \gev$ corresponding to the PIXIE sensitivity and the colored regions are excluded. The parameter sets producing the distortion of the order $|\Delta|\approx 3 \times 10^{-9}$ (corresponding to the expected magnitude in the conventional standard cosmology as discussed in the introduction section) are also shown to indicate the ultimate precision limit for the CMB spectral distortion measurements.  The other experimental constraints are adopted from~\cite{Alekhin:2015byh}.

Fig.~\ref{fig:EpsMchiPro} shows the exclusion plots on the plane of the DM mass ($m_\chi$) and the kinetic mixing ($\varepsilon^2$). The expected excluded regions from the CMB spectral distortion with a PIXIE-like sensitivity due to the elastic scattering between DM-proton (solid line) and those for the DM-electron (dashed line) scattering are shown with different colors representing different dark photon masses $m_{Z_d}$ ($\alpha_d$ = 0.1 is used for concreteness). We expect the momentum transfer is most efficient when two scattering particles are of the same mass and our figure indeed confirms that the bound from the spectral distortion becomes tightest when the DM mass is around the proton mass for the DM-protons scattering and around the electron mass for the DM-electrons scattering. 

%%%%%%%%%%%

\begin{figure}[t!]
\centering
\subfigure[DM-proton scattering]{\includegraphics[width=9cm]{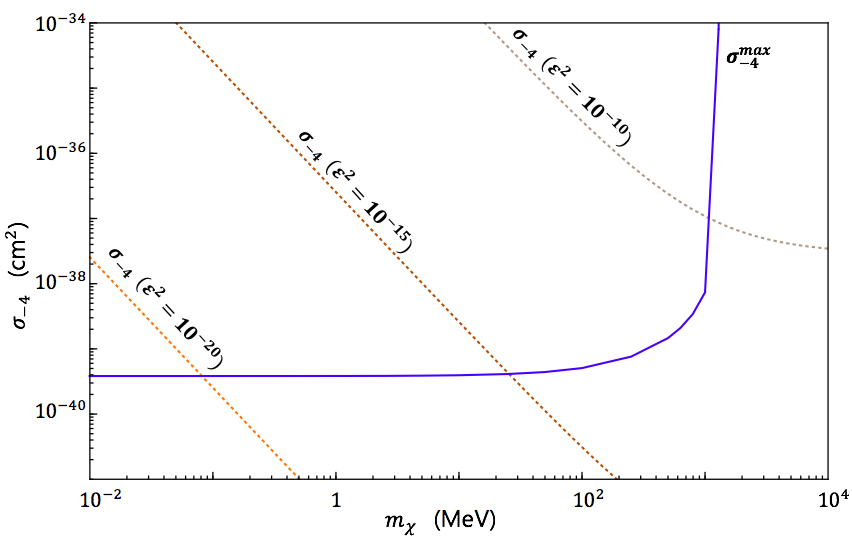}}
\subfigure[DM-electron scattering]{\includegraphics[width=9cm]{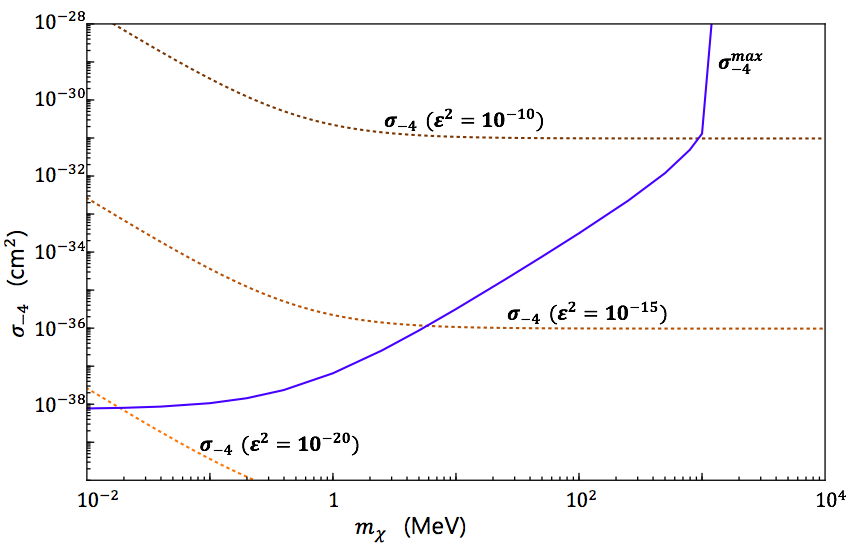}}
\caption{Expected upper bound on the momentum-transfer cross section $\sigma_{-4}$ with $\sigma = \sigma_{-4} v^{-4}$ (blue solid) for DM-protons (a), DM-electrons scattering (b).  Three dotted lines are the predictions from the light dark photon model with $m_{Z_d}\ll |\vec{k}|$ with $\varepsilon^2=10^{-20}, 10^{-15},$ and $10^{-10}$ respectively. Here we used $\alpha_d=0.1$ and  $m_{Z_d}=1$ eV. 
\label{fig:Bndnm4Pro}
}
\end{figure}

%%%%%%%%%%%

{An interesting feature is that the constraints due to the DM-proton interaction is stronger at $m_\chi \sim 100\mev$ than those due to the DM-electron interaction even though $\sigma_0^{max}(\chi-p)$ is approximately $100$ times larger than $\sigma_0^{max}(\chi-e)$.
This is because the cross section for DM-proton interaction is larger than that for DM-electron by $m_p^2/m_e^2$ as seen in Fig.~\ref{Bndn0}, thus the constraint becomes stronger compensating for the larger upper bound.

%%%%%%%%%%%%%%
\begin{figure}[t!]
\centering
\includegraphics[width=9cm]{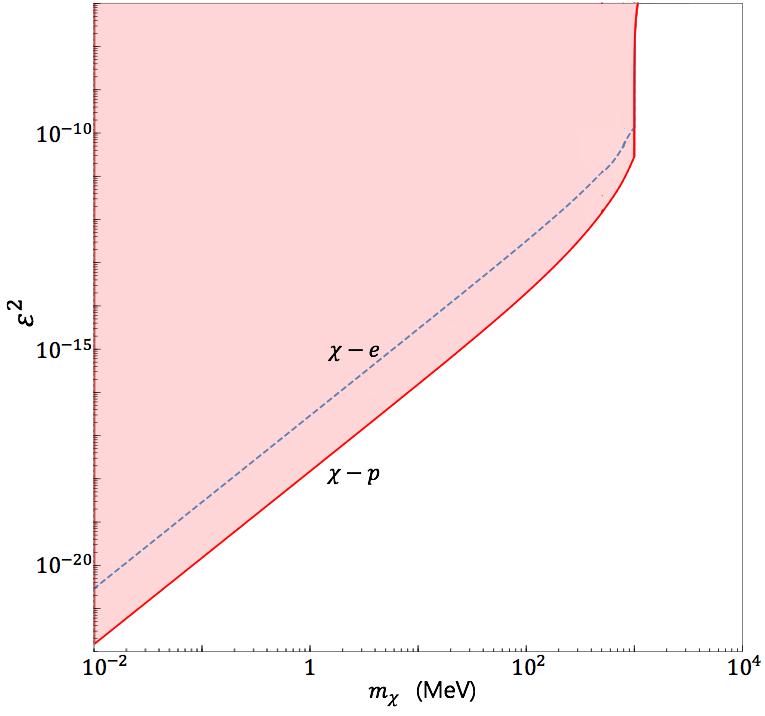}
\caption{The experimental bounds (colored regions are excluded) in terms of the DM mass ($m_\chi$) and the kinetic mixing ($\varepsilon$) for $m_{Z_d}\ll \kev$. The expected excluded regions from the CMB spectral distortion by PIXIE due to elastic scatterings between DM-proton (solid line) and DM-electron (dashed line) are shown with the dark photon mass $m_{Z_d}= 1\ev$ and $\alpha_d$ = 0.1 for concreteness.\label{EpsMchinm4}}
\end{figure}
%%%%%%%%%%%%%

Our discussions so far focused on the dark photon mass larger than the scale of the exchanged momentum $m_{Z_d}\gg |\vec{k}| $, where the velocity dependence in the momentum-transfer cross section disappears at the leading order. We briefly discuss, before concluding our study, the opposite limit for a small dark photon mass, where the cross section behaves as $\sigma \sim v^{-4}$. 

For $m_{Z_d}\ll |\vec{k}|$,  the differential cross section becomes
\begin{equation}
 \frac{d\sigma}{d\Omega}\simeq\frac{4c_W^2\varepsilon^2\alpha\alpha_Dm_\chi^2 m_b^2}{(m_\chi+m_b)^2}\frac{1}{(2\vec{k}^2(1-\cos\theta)+m_{Z_d}^2)^2} ,
 \end{equation}
and the corresponding momentum transfer cross section is
\dis{
\sigma_{tr}&\simeq2\pi c_W^2\varepsilon^2\alpha\alpha_D\frac{m_b^2m_\chi^2}{(m_b+m_\chi)^2\vec{k}^4}\left[\ln{\left(\frac{4\vec{k}^2}{m_{Z_d}^2}\right)}-1\right],\\
&\simeq 2\pi c_W^2\varepsilon^2\alpha\alpha_D\frac{(m_b+m_\chi)^2}{m_b^2m_\chi^2v^4}\left[\ln{\left(\frac{4\times10^4\, eV^2}{m_{Z_d}^2}\right)}-1\right],\\
&\equiv \sigma_{-4} v^{-4}.
\label{eq:largek}
}
In the second line we used the relation between $|\vec{k}|$ and $v$, and used the approximation that the logarithmic term does not change much during the epoch of our interest for $10^6\lesssim z\lesssim 10^4$ (we thus used $|\vec{k}|=100 \ev$, a typical momentum scale around $z\sim 10^6$).

While the DM decoupling epoch can be approximated by the step function for $n=0$,  the DM kinetic decoupling is far from instantaneous transition for a light $m_{Z_d}$ where $n=-4$. Therefore instead of using the step-function approximation as done in~\cite{Ali-Haimoud:2015pwa}, we here solve Eqs.~(\ref{tempeq}) and~(\ref{derivrho}) numerically to obtain the upper bound on the momentum transfer cross section.  The corresponding bound is shown in Fig.~\ref{fig:Bndnm4Pro}~\footnote{For $n\leq -2$, the thermal decoupling is gradual and the Maxwell-Boltzmann distribution would not be a good approximation ~\cite{Ali-Haimoud:2015pwa}. We need, in this case, a more rigorous treatment by solving the Boltzmann equation in the phase space and defer it to our future work.}. We can see the bound has little dependence on the DM mass, which can be expected from  \eq{derivrho} characterizing the magnitude of the spectral distortion. For a light DM, $r_{\chi b} N_\chi$ in \eq{derivrho} is independent of DM mass because $\Gamma_{\chi b} \propto m_\chi$ and $N_\chi/N^{tot}_b \sim 3(\gev/m_\chi) $. The mass dependence shows up for a larger DM mass $m_\chi \gtrsim m_b$ where $\Gamma_{\chi b} \propto 1/m_\chi$ and thus $r_{\chi b} N_\chi \propto 1/m_\chi^2$, before the distortion signals become too small to be detected for $m_\chi \gtrsim  1.3 \gev$. Also note the bounds have little dependence on the dark photon mass $m_{Z_d}$ because the cross section only depends logarithmically on $m_{Z_d}$. This is reasonable because the dark photon propagator $~1/(k^2-m_{Z_d}^2)$ has a small dependence on  $m_{Z_d}$ when $m_{Z_d}\ll k$.
Fig.~\ref{EpsMchinm4} shows the expected constraints on $m_\chi$ and $\varepsilon^2$ from the DM-proton (solid lines) scattering and the DM-electron scattering (dashed lines) for $m_{Z_d}= 1\ev$ and $\alpha_d$ = 0.1.

\section{Conclusion}
\label{Conclusion}
We have explored the possibilities to probe the dark sector where the hidden gauge boson kinetically mixes with the SM from the CMB spectral distortion. 
The momentum transfer between baryon-photon plasma and DM can extract energy from CMB and distort their spectra. We studied the effects in the dark photon model as a concrete example beyond the SM. In particular, we focused on a relatively light (sub-GeV) dark photon for detectable distortions in the CMB spectra, and studied the expected bounds from the future experiments such as PIXIE. We pointed out the different velocity dependence of the cross section for a different dark photon mass and we presented the bounds on the dark photon model in the regimes for large and small masses of dark photon corresponding to $n=0$ and $-4$ (the power of the cross section $\propto v^n$) respectively.

  While the stringent bounds already exist on the dark photon model, in particular, from the collider experiments, we illustrated that the astrophysical observables can also give the compelling limits on the dark photon parameters totally independent from those coming from the particle physics experiments. Our new constraints from the CMB spectral distortion are comparable with those already existing constraints at $m_{Z_d}= 10 \rm MeV$. More specifically, we found the CMB spectral distortion observables can give the tight bounds, for $m_{Z_d}\gg \kev$ (which corresponds to $n=0$), when $m_\chi\sim m_p (\rm GeV)$ for $\chi-p$ scattering and when $m_\chi\sim m_e (\rm MeV)$ for $\chi-e$ scattering. It can be understood by the fact that momentum transfer is maximized when the scattering particles have comparable masses. The DM-electron scattering can give the tighter bounds than that from DM-proton scattering for a lighter dark matter mass range as illustrated in Fig.~\ref{fig:EpsMchiPro}.  For $m_{Z_d}\ll \kev$ (which corresponds to $n=-4$), in contrast, $\chi-p$ scattering gives stronger constraints than $\chi-e$ scattering for the dark matter mass range considered in our analysis. This is because, as Fig.~\ref{fig:Bndnm4Pro} illustrates, the upper bound on the momentum-transfer cross section of $\chi-p$ scattering is always stronger than $\chi-e$ scattering.

We leave the study for a more general dark photon mass range taking account of the collisional Boltzmann equations without assuming the Maxwell-Boltzmann distribution for our future work.

\section*{Acknowledgment}
K.-Y.C. was supported by the National Research Foundation of Korea(NRF) grant funded by the Korea government(MEST) (NRF-2016R1A2B4012302). KK and I.Park were supported by Institute for Basic Science (IBS-R018-D1). K.-Y.C. appreciates Asia Pacific Center for Theoretical Physics for the support to the Focus Research Program.

%%%%%%%%%%%%%%%%%%%%%%%%%%%%%%%%%%%%%%%%%%%%%%%%%%%%%%%%%%%%

\section*{References}
%\bibliography{mybibfile}

\begin{thebibliography}{99}
%%%%%%%%%%%%%%%%%%%%%%%%%%%%%%%%%%%%%%%%%%%%%%%%%%%%%%%%%%%%


%\cite{Sunyaev:1970er}
\bibitem{Sunyaev:1970er} 
  R.~A.~Sunyaev and Y.~B.~Zeldovich,
  %``The Interaction of matter and radiation in the hot model of the universe,''
  Astrophys.\ Space Sci.\  {\bf 7}, 20 (1970).
  %%CITATION = APSSB,7,20;%%
  %165 citations counted in INSPIRE as of 15 Dec 2016

  %\cite{Zeldovich:1969ff}
\bibitem{Zeldovich:1969ff} 
  Y.~B.~Zeldovich and R.~A.~Sunyaev,
  %``The Interaction of Matter and Radiation in a Hot-Model Universe,''
  Astrophys.\ Space Sci.\  {\bf 4}, 301 (1969).
  doi:10.1007/BF00661821
  %%CITATION = doi:10.1007/BF00661821;%%
  %341 citations counted in INSPIRE as of 15 Dec 2016


%\cite{Fixsen:1996nj}
\bibitem{Fixsen:1996nj}
  D.~J.~Fixsen, E.~S.~Cheng, J.~M.~Gales, J.~C.~Mather, R.~A.~Shafer and E.~L.~Wright,
  %``The Cosmic Microwave Background spectrum from the full COBE FIRAS data set,''
  Astrophys.\ J.\  {\bf 473} (1996) 576
  doi:10.1086/178173
  [astro-ph/9605054].
  %%CITATION = doi:10.1086/178173;%%


%\cite{Kogut:2011xw}
\bibitem{Kogut:2011xw}
  A.~Kogut {\it et al.},
  %``The Primordial Inflation Explorer (PIXIE): A Nulling Polarimeter for Cosmic Microwave Background Observations,''
  JCAP {\bf 1107} (2011) 025
  doi:10.1088/1475-7516/2011/07/025
  [arXiv:1105.2044 [astro-ph.CO]].
  %%CITATION = doi:10.1088/1475-7516/2011/07/025;%%

%\cite{Chluba:2011hw}
\bibitem{Chluba:2011hw}
  J.~Chluba and R.~A.~Sunyaev,
  %``The evolution of CMB spectral distortions in the early Universe,''
  Mon.\ Not.\ Roy.\ Astron.\ Soc.\  {\bf 419} (2012) 1294
  doi:10.1111/j.1365-2966.2011.19786.x
  [arXiv:1109.6552 [astro-ph.CO]].
  %%CITATION = doi:10.1111/j.1365-2966.2011.19786.x;%%

%\cite{Hu:1993gc}
\bibitem{Hu:1993gc}
  W.~Hu and J.~Silk,
  %``Thermalization constraints and spectral distortions for massive unstable relic particles,''
  Phys.\ Rev.\ Lett.\  {\bf 70} (1993) 2661.
  doi:10.1103/PhysRevLett.70.2661
  %%CITATION = doi:10.1103/PhysRevLett.70.2661;%%
  
  %\cite{Sarkar:1984tt}
\bibitem{Sarkar:1984tt} 
  S.~Sarkar and A.~M.~Cooper-Sarkar,
  %``Cosmological and experimental constraints on the tau neutrino,''
  Phys.\ Lett.\  {\bf 148B}, 347 (1984).
  doi:10.1016/0370-2693(84)90101-1
  %%CITATION = doi:10.1016/0370-2693(84)90101-1;%%
  %99 citations counted in INSPIRE as of 20 Apr 2017
  
  %\cite{Carr:2009jm}
\bibitem{Carr:2009jm}
  B.~J.~Carr, K.~Kohri, Y.~Sendouda and J.~Yokoyama,
  %``New cosmological constraints on primordial black holes,''
  Phys.\ Rev.\ D {\bf 81} (2010) 104019
  doi:10.1103/PhysRevD.81.104019
  [arXiv:0912.5297 [astro-ph.CO]].
  %%CITATION = doi:10.1103/PhysRevD.81.104019;%%


%\cite{McDonald:2000bk}
\bibitem{McDonald:2000bk} 
  P.~McDonald, R.~J.~Scherrer and T.~P.~Walker,
  %``Cosmic microwave background constraint on residual annihilations of relic particles,''
  Phys.\ Rev.\ D {\bf 63}, 023001 (2001)
  doi:10.1103/PhysRevD.63.023001
  [astro-ph/0008134].
  %%CITATION = doi:10.1103/PhysRevD.63.023001;%%
  %44 citations counted in INSPIRE as of 13 Jan 2017

%\cite{Padmanabhan:2005es}
\bibitem{Padmanabhan:2005es}
  N.~Padmanabhan and D.~P.~Finkbeiner,
  %``Detecting dark matter annihilation with CMB polarization: Signatures and experimental prospects,''
  Phys.\ Rev.\ D {\bf 72} (2005) 023508
  doi:10.1103/PhysRevD.72.023508
  [astro-ph/0503486].
  %%CITATION = doi:10.1103/PhysRevD.72.023508;%%
  %165 citations counted in INSPIRE as of 26 Dec 2016


%\cite{Sunyaev:1970}
\bibitem{Sunyaev:1970} 
  R.~A.~Sunyaev and Y.~B.~Zeldovich,
  %``Small scale entropy and adiabatic density perturbations: Antimatter in the universe.,''
  Astrophys.\ Space Sci.\  {\bf 9}, 368 (1970).
  %%CITATION = APSSB,7,20;%%
  %56 citations counted 20 Apr 2017

%\cite{Chluba:2012gq}
\bibitem{Chluba:2012gq} 
  J.~Chluba, R.~Khatri and R.~A.~Sunyaev,
  %``CMB at 2x2 order: The dissipation of primordial acoustic waves and the observable part of the associated energy release,''
  Mon.\ Not.\ Roy.\ Astron.\ Soc.\  {\bf 425}, 1129 (2012)
  doi:10.1111/j.1365-2966.2012.21474.x
  [arXiv:1202.0057 [astro-ph.CO]].
  %%CITATION = doi:10.1111/j.1365-2966.2012.21474.x;%%
  %89 citations counted in INSPIRE as of 20 Apr 2017


%\cite{Silk:1967kq}
\bibitem{Silk:1967kq}
  J.~Silk,
  %``Cosmic black body radiation and galaxy formation,''
  Astrophys.\ J.\  {\bf 151} (1968) 459.
  doi:10.1086/149449
  %%CITATION = doi:10.1086/149449;%%
  %464 citations counted in INSPIRE as of 26 Dec 2016




%\cite{Khatri:2011aj}
\bibitem{Khatri:2011aj}
  R.~Khatri, R.~A.~Sunyaev and J.~Chluba,
  %``Does Bose-Einstein condensation of CMB photons cancel \mu distortions created by dissipation of sound waves in the early Universe?,''
  Astron.\ Astrophys.\  {\bf 540} (2012) A124
  doi:10.1051/0004-6361/201118194
  [arXiv:1110.0475 [astro-ph.CO]].
  %%CITATION = doi:10.1051/0004-6361/201118194;%%
  
  %\cite{Pajer:2013oca}
\bibitem{Pajer:2013oca}
  E.~Pajer and M.~Zaldarriaga,
  %``A hydrodynamical approach to CMB $\mu$-distortion from primordial perturbations,''
  JCAP {\bf 1302} (2013) 036
  doi:10.1088/1475-7516/2013/02/036
  [arXiv:1206.4479 [astro-ph.CO]].
  %%CITATION = doi:10.1088/1475-7516/2013/02/036;%%



  %\cite{Tashiro:2014tsa}
\bibitem{tashiro2014} 
  H.~Tashiro, K.~Kadota and J.~Silk,
  %``Effects of dark matter-baryon scattering on redshifted 21 cm signals,''
  Phys.\ Rev.\ D {\bf 90}, no. 8, 083522 (2014)
  doi:10.1103/PhysRevD.90.083522
  [arXiv:1408.2571 [astro-ph.CO]].
  %%CITATION = doi:10.1103/PhysRevD.90.083522;%%
  %5 citations counted in INSPIRE as of 27 Nov 2016
  
%\cite{Ali-Haimoud:2015pwa}
\bibitem{Ali-Haimoud:2015pwa}
  Y.~Ali-HaÂmoud, J.~Chluba and M.~Kamionkowski,
  %``Constraints on Dark Matter Interactions with Standard Model Particles from Cosmic Microwave Background Spectral Distortions,''
  Phys.\ Rev.\ Lett.\  {\bf 115} (2015) no.7,  071304
  doi:10.1103/PhysRevLett.115.071304
  [arXiv:1506.04745 [astro-ph.CO]].
  %%CITATION = doi:10.1103/PhysRevLett.115.071304;%%


%\cite{Okun:1982xi}
\bibitem{Okun:1982xi} 
  L.~B.~Okun,
  %``Limits Of Electrodynamics: Paraphotons?,''
  Sov.\ Phys.\ JETP {\bf 56}, 502 (1982)
  [Zh.\ Eksp.\ Teor.\ Fiz.\  {\bf 83}, 892 (1982)].
  %%CITATION = SPHJA,56,502;%%
  %230 citations counted in INSPIRE as of 15 Dec 2016

%\cite{Holdom:1985ag}
\bibitem{Holdom:1985ag} 
  B.~Holdom,
  %``Two U(1)'s and Epsilon Charge Shifts,''
  Phys.\ Lett.\  {\bf 166B}, 196 (1986).
  doi:10.1016/0370-2693(86)91377-8
  %%CITATION = doi:10.1016/0370-2693(86)91377-8;%%
  %1057 citations counted in INSPIRE as of 15 Dec 2016

%\cite{Davoudiasl:2012ag}
\bibitem{Davoudiasl:2012ag}
  H.~Davoudiasl, H.~S.~Lee and W.~J.~Marciano,
  %``'Dark' Z implications for Parity Violation, Rare Meson Decays, and Higgs Physics,''
  Phys.\ Rev.\ D {\bf 85} (2012) 115019
  doi:10.1103/PhysRevD.85.115019
  [arXiv:1203.2947 [hep-ph]].
  %%CITATION = doi:10.1103/PhysRevD.85.115019;%%


    %\cite{Essig:2013lka}
\bibitem{Essig:2013lka}
  R.~Essig {\it et al.},
  %``Working Group Report: New Light Weakly Coupled Particles,''
  arXiv:1311.0029 [hep-ph].
  %%CITATION = ARXIV:1311.0029;%%
  %245 citations counted in INSPIRE as of 26 Dec 2016
  
  
  %\cite{Adare:2014mgk}
\bibitem{Adare:2014mgk} 
  A.~Adare {\it et al.} [PHENIX Collaboration],
  %``Search for dark photons from neutral meson decays in $p + p$ and $d$ + Au collisions at $\sqrt{s_{NN}} =$ 200 GeV,''
  Phys.\ Rev.\ C {\bf 91}, no. 3, 031901 (2015)
  doi:10.1103/PhysRevC.91.031901
  [arXiv:1409.0851 [nucl-ex]].
  %%CITATION = doi:10.1103/PhysRevC.91.031901;%%
  %35 citations counted in INSPIRE as of 16 Dec 2016

  
  %\cite{Goudzovski:2014rwa}
\bibitem{Goudzovski:2014rwa} 
  E.~Goudzovski [NA48/2 Collaboration],
  %``Search for the dark photon in $?^0$ decays by the NA48/2 experiment at CERN,''
  EPJ Web Conf.\  {\bf 96}, 01017 (2015)
  doi:10.1051/epjconf/20159601017
  [arXiv:1412.8053 [hep-ex]].
  %%CITATION = doi:10.1051/epjconf/20159601017;%%
  %13 citations counted in INSPIRE as of 16 Dec 2016
  
  %\cite{Lees:2015rxq}
\bibitem{Lees:2015rxq} 
  J.~P.~Lees {\it et al.} [BaBar Collaboration],
  %``Search for Long-Lived Particles in $e^+e^-$ Collisions,''
  Phys.\ Rev.\ Lett.\  {\bf 114}, no. 17, 171801 (2015)
  doi:10.1103/PhysRevLett.114.171801
  [arXiv:1502.02580 [hep-ex]].
  %%CITATION = doi:10.1103/PhysRevLett.114.171801;%%
  %9 citations counted in INSPIRE as of 16 Dec 2016


%\cite{Alekhin:2015byh}
\bibitem{Alekhin:2015byh} 
  S.~Alekhin {\it et al.},
  %``A facility to Search for Hidden Particles at the CERN SPS: the SHiP physics case,''
  Rept.\ Prog.\ Phys.\  {\bf 79}, no. 12, 124201 (2016)
  doi:10.1088/0034-4885/79/12/124201
  [arXiv:1504.04855 [hep-ph]].
  %%CITATION = doi:10.1088/0034-4885/79/12/124201;%%
  %129 citations counted in INSPIRE as of 15 Dec 2016

%\cite{Dvorkin:2013cea}
\bibitem{Dvorkin:2013cea}
  C.~Dvorkin, K.~Blum and M.~Kamionkowski,
  %``Constraining Dark Matter-Baryon Scattering with Linear Cosmology,''
  Phys.\ Rev.\ D {\bf 89} (2014) no.2,  023519
  doi:10.1103/PhysRevD.89.023519
  [arXiv:1311.2937 [astro-ph.CO]].
  %%CITATION = doi:10.1103/PhysRevD.89.023519;%%



  %\cite{Berger:2016vxi}
\bibitem{Berger:2016vxi}
  J.~Berger, K.~Jedamzik and D.~G.~E.~Walker,
  %``Cosmological Constraints on Decoupled Dark Photons and Dark Higgs,''
  arXiv:1605.07195 [hep-ph].
  %%CITATION = ARXIV:1605.07195;%%



%\cite{Raby:1987nb}
\bibitem{raby1987} 
  S.~A.~Raby and G.~West,
  %``A Simple Solution to the Solar Neutrino and Missing Mass Problems,''
  Nucl.\ Phys.\ B {\bf 292}, 793 (1987).
  doi:10.1016/0550-3213(87)90671-7
  %%CITATION = doi:10.1016/0550-3213(87)90671-7;%%
  %57 citations counted in INSPIRE as of 28 Nov 2016

    
  
  \end{thebibliography}

\end{document}